\documentclass[conference]{IEEEtran}
\IEEEoverridecommandlockouts
\usepackage{cite}
\usepackage{amsmath,amssymb,amsfonts}
\usepackage{algorithmic}
\usepackage{graphicx}
\usepackage{textcomp}
\usepackage{xcolor}

\usepackage{bbm}
\usepackage{dsfont}
\usepackage{bbold}
\usepackage{tabularray}
\usepackage{stfloats}
\usepackage{siunitx}
\usepackage{mathtools}

\usepackage{todonotes}

\newcommand{\bfd}{\ensuremath{\mathbf{d}}}

\newcommand{\bfh}{\ensuremath{\mathbf{h}}}

\newcommand{\bfv}{\ensuremath{\mathbf{v}}}
\newcommand{\bfw}{\ensuremath{\mathbf{w}}}
\newcommand{\bfx}{\ensuremath{\mathbf{x}}}

\newcommand{\bfz}{\ensuremath{\mathbf{z}}}

\newcommand{\bfB}{\ensuremath{\mathbf{B}}}

\newcommand{\bfF}{\ensuremath{\mathbf{F}}}

\newcommand{\bfH}{\ensuremath{\mathbf{H}}}
\newcommand{\bfI}{\ensuremath{\mathbf{I}}}

\newcommand{\bfQ}{\ensuremath{\mathbf{Q}}}
\newcommand{\bfR}{\ensuremath{\mathbf{R}}}

\newcommand{\bfV}{\ensuremath{\mathbf{V}}}
\newcommand{\bfW}{\ensuremath{\mathbf{W}}}
\newcommand{\bfX}{\ensuremath{\mathbf{X}}}

\newcommand{\bfZ}{\ensuremath{\mathbf{Z}}}

\newcommand{\bbfZ}{\ensuremath{\bar{\mathbf{Z}}}}

\newcommand{\bbN}{\ensuremath{\mathbb{N}}}

\newcommand{\bbR}{\ensuremath{\mathbb{R}}}

\newcommand{\mcT}{\ensuremath{\mathcal{T}}}

\DeclareMathOperator*{\mean}{\mathbb{E}}				% Expectation operator
\DeclareMathOperator*{\cov}{cov}			% Covariance operator
\DeclareMathOperator*{\var}{var}			% Variance operator
\DeclareMathOperator*{\rank}{rank}			% Rank operator
\DeclareMathOperator*{\diag}{diag}			% Vectorization operator
\newcommand{\indfun}{\ensuremath{\mathds{1}}} % Indicator function requires \usepackage{dsfont}
\newcommand{\T}{\ensuremath{\text{T}}} % Transpose
\newcommand{\Ts}{\ensuremath{T_{\mathrm{s}}}}

\newcommand{\bfalpha}{\ensuremath{\boldsymbol{\alpha}}}

\newcommand{\bfmu}{\ensuremath{\boldsymbol{\mu}}}

\newcommand{\bftheta}{\ensuremath{\boldsymbol{\theta}}}

\newcommand{\bfnu}{\ensuremath{\boldsymbol{\nu}}}

\newcommand{\bfGamma}{\ensuremath{\boldsymbol{\Gamma}}}

\newcommand{\bfvarphi}{\ensuremath{\boldsymbol{\varphi}}}

\newcommand{\bfPhi}{\ensuremath{\boldsymbol{\Phi}}}
\newcommand{\bfepsilon}{\ensuremath{\boldsymbol{\epsilon}}}

\newcommand{\bfzeta}{\ensuremath{\boldsymbol{\zeta}}}

\RequirePackage{xspace}

\def\bfI{\mathbf I}
\def\bfV{\mathbf V}
\def\bfW{\mathbf W}
\def\bfX{\mathbf X}

\def\bfZ{\mathbf Z}

\def\real{\mathbb{R}}

\def\bfGamma{\mathbf \Gamma}

\def\calA{\mathcal{A}}

\def\calE{\mathcal{E}}
\def\calI{\mathcal{I}}
\def\calO{\mathcal{O}}
\def\calR{\mathcal{R}}

\def\bfnul{\mathbf{0}}

\def\T{^\intercal}

\def\anni{\mathsf{am}}
\def\rank{\mathsf{rank}}

\def\bbfZ{\bar{\bfZ}}
\def\tbfZ{\tilde{\bfZ}}

\def\aO{n_{\mathsf{a}\raisebox{-0.45mm}{\scriptsize$\calO$}}}

\def\re{n_\calE}

\def\diag{\mathsf{diag}}
\def\vertcat{\mathsf{vertcat}}

\newcommand{\bfza}{\bfz_{\mathrm{a}}}
\newcommand{\bfthetaa}{\bftheta_{\mathrm{a}}}

\begin{document}

\title{Identification of Clock Ensemble Noise Parameters Using Differential Measurement Analysis
\thanks{This work was created with the support of the project "R\&D of Technologies for Advanced Digitization in the Pilsen Metropolitan Area (DigiTech)" No.: CZ.02.01.01/00/23\_021/0008436 co-financed by the European Union.}

%% can be used in arXiv version
% \thanks{\copyright 2026 IEEE. Personal use of this material is permitted. Permission from IEEE must be obtained for all other uses, in any current or future media, including reprinting/republishing this material for advertising or promotional purposes, creating new collective works, for resale or redistribution to servers or lists, or reuse of any copyrighted component of this work in other works. This work has been published at European Frequency and Time Forum (EFTF 2026). DOI: xyz}

}

% \author{
%     \IEEEauthorblockN{Jindrich Dunik\IEEEauthorrefmark{1}, Ladislav Kral \IEEEauthorrefmark{1}, Ivo Puncochar\IEEEauthorrefmark{1}, Oliver Kost\IEEEauthorrefmark{1}, Ondrej Daniel\IEEEauthorrefmark{2},Simona Circiu\IEEEauthorrefmark{3}, Bernardino Quaranta\IEEEauthorrefmark{3} }
%     \IEEEauthorblockA{\IEEEauthorrefmark{1}University of West Bohemia, Pilsen, Czech Republic  \\ \{dunikj,ladkral,ivop,kost\}@fav.zcu.cz}
%     \IEEEauthorblockA{\IEEEauthorrefmark{2}HULD s.r.o., Prague, Czech Republic
%     \\\{2, 3\}@def.com}
%     \IEEEauthorblockA{\IEEEauthorrefmark{2}ESA/ESTEC, Noordwijk, The Netherlands
%     \\\{2, 3\}@def.com}    
% }

\author{\IEEEauthorblockN{Jind\v{r}ich Dun\'{i}k, Ladislav Kr\'{a}l,\\ Ivo Pun\v{c}och\'{a}\v{r}, Oliver Kost}
\IEEEauthorblockA{\textit{University of West Bohemia} \\
Pilsen, Czech Republic\\
 \{dunikj,ladkral,ivop,kost\}@fav.zcu.cz}
\and
\IEEEauthorblockN{Ond\v{r}ej Daniel}
\IEEEauthorblockA{\textit{HULD s.r.o.} \\
Prague, Czech Republic \\
ondrej.daniel@huld.io}
\and
\IEEEauthorblockN{Simona Circiu, Bernardino Quaranta}
\IEEEauthorblockA{\textit{ESA/ESTEC, Noordwijk} \\
Noordwijk, The Netherlands \\
\{bernardino.quaranta,simona.circiu\}@esa.int}
}

% Adding copyright information to the first page
\IEEEoverridecommandlockouts
\IEEEpubid{\makebox[\columnwidth]{979-8-3315-8115-2/26/\$31.00~\copyright2026 IEEE \hfill} \hspace{\columnsep}\makebox[\columnwidth]{ }}

\maketitle

\begin{abstract}
The paper addresses the critical problem of identifying unknown parameters of an atomic clock ensemble. The ensemble model is considered as a set of individual clock models, where each clock is described by a second-order linear stochastic state-space model. The paper presents identification procedure for model unknown parameters based solely on the availability of differential measurements - that is, the measured pairwise phase differences between a designated pivot clock and all other clocks within the ensemble. Specifically, each clock model is defined by the following set of unknown parameters: the variances characterizing the white frequency noise and random walk frequency noise, the drift, and the (co)variance of the measurement noise. Two distinct identification methods are designed to estimate the unknown clock model parameters. The accuracy of the identified sets of parameters are demonstrated on a simulation scenario/real data combining atomic H-maser (AHM) clocks.%, where dataset was acquired at the ESA ESTEC Time Laboratory as part of the ESA project NAVISP-EL1-056.
\end{abstract}


\section{Introduction}
Time plays a fundamental role in modern science, technology, and society, underpinning applications ranging from navigation and telecommunications to financial systems and distributed networks \cite{Allan1997,Galleani2010:IEEECon}. High-quality time scales are typically generated by an ensemble of atomic clocks, where the aggregation process reduces the impact of individual clock frequency instability and improves overall performance. Such clock ensembles form the basis of internationally recognized time scales and provide robustness against individual clock failures.

One of the most promising directions for improving the quality of time scales in terms of stability, integrity, and availability is the adoption of model-based approach. This enables the incorporation of prior knowledge about clock behavior and noise characteristics into the estimation process. In particular, advanced time combination and fault detection algorithms can significantly benefit from accurate stochastic models, allowing for more reliable weighting strategies and timely identification of anomalous clocks within the ensemble.

A critical challenge in this context is the identification of unknown noise parameters of the clock ensemble. This problem has been extensively studied in the literature, with foundational contributions addressing stochastic clock modeling and noise characterization methods. Notable works include those focused on Allan variance and its variants, the maximum likelihood method as well as methods for estimating power-law noise parameters and covariance structures in clock ensembles \cite{Tryon1983,Barnes1983a,Premoli1993:IEEETIM,Tavella1993:Metrologia,Vernotte2016}. Despite these efforts, accurate parameter identification remains challenging, especially under practical constraints such as limited data availability or indirect observations.

This paper addresses the problem of identifying noise parameters in a clock ensemble when only differential measurements between a selected pivot clock and the remaining clocks are available. The main objective is to develop an effective identification method tailored to this measurement configuration, enabling reliable estimation of ensemble noise characteristics under realistic observational constraints.

\section{A clock and ensemble model definition} \label{sec:ensembleDef}
\IEEEpubidadjcol %MUST be placed anywhere in the second column
It is assumed that the $i$-th clock of an ensemble can be described at a time step $k\in\mcT=\{0,1,2\ldots,N\}$ by a second-order discrete-time state-space model\footnote{Notation: A symbol $x^{(i)}$ denotes a parameter or a variable belonging to the $i$-th clock, $x_{i,k}$ is the $i$-th component of vector $\bfx$ in the time instant $k$, $x_{ij}$ is the $(i,j)$-th element of the matrix $\bfX$, $\bfI_n \in \bbR^{n \times n}$ is an $n$-dimensional identity matrix, $\bfx_{i:j} = [x_{i}, x_{i+1}, \ldots, x_{j} ]^{\T}$ is a sequence of $x$ from step $i$ to step $j$, $\indfun_{m}\in \bbR^m$ is a vector of ones, $\mathbf{0}_{m\times n} \in \bbR^{m \times n}$ is the zero matrix, $\hat{x}$ is an estimate of $x$, $\otimes$ is the Kronecker product, $\vertcat$ is a vertical concatenating operation, $\diag$ is a block diagonal concatenation operator, $\mean$ is the expectation operator, and $\var$, $\cov$ are operators of variance and covariance. The symbols $\bbR^{+}$ and $\bbR^{++}$ denote real non-negative and real positive numbers, respectively.}
\begin{align}
    \bfx_{k+1}^{(i)} & = \bfF_1 \bfx_k^{(i)} + \bfw_k^{(i)},  \label{eq:clk1}\\
    \Delta_k^{(i)} & = \bfh_{1}^{\T} \bfx_k^{(i)},  \label{eq:clk2}
\end{align}
where $\bfx_k^{(i)} = [x_{1,k}^{(i)}, \ x_{2,k}^{(i)}]^{\T} \in \bbR^2$ is a state consisting of a time deviation $x_{1,k}^{(i)}\in\bbR$ and a random walk process $x_{2,k}^{(i)}\in \bbR$, $\bfw_k^{(i)}\in\bbR^{2}$ is a white Gaussian state noise with mean $\bfmu^{(i)}\in\bbR^{2}$ and covariance matrix $\bfQ^{(i)}\in\bbR^{2 \times 2}$, and $\Delta_{k}^{(i)}\in\bbR$ is a time deviation. The matrices $\bfF_{1}$, $\bfQ^{(i)}$ and vectors $\bfh_{1}$, $\bfmu^{(i)}$ are given as
\begin{equation}
    \begin{aligned}
        \bfF_1 =
    \begin{bmatrix} 
        1 & \Ts \\
        0 & 1 
    \end{bmatrix}&,\ 
    \bfh_{1}^{\T} = 
    \begin{bmatrix} 
        1 & 0 
    \end{bmatrix},\ 
    \bfmu^{(i)} = d^{(i)} 
    \begin{bmatrix} 
        \Ts^2/2 \\
        \Ts 
    \end{bmatrix},\\
    \bfQ^{(i)}& = 
    \begin{bmatrix} 
        q_1^{(i)} \Ts + q_2^{(i)} \Ts^3/3 & q_2^{(i)} \Ts^2/2 \\ 
            q_2^{(i)} \Ts^2/2 & q_2^{(i)} \Ts
    \end{bmatrix}, 
    \end{aligned}
\end{equation}
where $\Ts\in \bbR^{++}$ is a sampling period in seconds, $d^{(i)} \in \bbR$ is a frequency drift, and $q_1^{(i)} \in \bbR^{++}$ and  $q_2^{(i)}\in \bbR^{+++}$ denote intensities of original continuous-time stochastic processes that represent the white noise frequency modulation and random walk frequency modulation, respectively.

The model of ensemble of $n$ clocks can be defined as
\begin{align}
    \bfx_{k+1} &= \bfF \bfx_k + \bfw_k, \label{eq:ensemble_state} \\
    \bfz_k &=  \bfH \bfx_k + \bfv_k,\label{eq:ensemble_meas}
\end{align}
where $\bfx_{k} \in \bbR^{2n}$ is an ensemble state, $\bfw_{k} \in \bbR^{2n}$ is an ensemble state noise with mean $\bfmu\in\bbR^{2n}$ and covariance matrix $\bfQ\in\bbR^{2n \times 2n}$, $\bfz_k \in \bbR^{n_z}$ is a measurement containing $n_z = n-1$ differences of time deviations of a pivot clock and all other clocks, and $\bfv_{k} \in \bbR^{n_z}$ is a white Gaussian measurement noise with zero mean and covariance matrix $\bfR \in \bbR^{n_z \times n_z}$. The vectors and matrices are given as
\begin{align}
    \bfx_k &= \vertcat \left( \bfx_k^{(1)}, \bfx_k^{(2)}, \dots, \bfx_k^{(n)} \right),\\
    \bfw_k &= \vertcat \left( \bfw_k^{(1)}, \bfw_k^{(2)}, \dots, \bfw_k^{(n)} \right),\\
    \bfz_k & = [z_{1,k}, z_{2,k},\ldots, z_{n_z,k}]^{\T}, \\
    \bfF &= \bfI_{n} \otimes \bfF_{1}, \quad \bfH = \bfV \left( \bfI_{n} \otimes \bfh_{1}^{\T} \right),\\
    \bfmu & = \vertcat \left( \bfmu^{(1)}, \bfmu^{(2)}, \dots, \bfmu^{(n)} \right),\\
    \bfQ &= \diag \left( \bfQ^{(1)}, \bfQ^{(2)}, \dots, \bfQ^{(n)} \right),
\end{align}  
where $\bfV = [-\mathds{1}_{n_z}, \bfI_{n_z}]\in\bbR^{n_z\times n}$ is a matrix that specifies the first clock in the ensemble to be the pivot clock.

\section{Problem Formulation} \label{sec:ProblemFormualtion}
Let's define a vector of unknown model parameters as
\begin{multline}
     \bftheta = \left[ q_1^{(1)}, q_1^{(2)}, \dots, q_1^{(n)}, q_2^{(1)}, q_2^{(2)}, \dots, q_2^{(n)}, \right.\\
     \left. d^{(1)}, d^{(2)},  \dots, d^{(n)}, r_{11}, r_{12}, \ldots, r_{n_{z}n_{z}}\right]^{\T},
\end{multline}
where $\bftheta \in \bbR^{n_\theta}$ with $n_\theta = n(n+5)/2$ denoting the number of the  unknown parameters  and $r_{ij}$, $i \leq j$ is an element of the upper triangular part of the matrix $\bfR$.

\textbf{Goal:} Given an available sequence of measurements  $\bfz_{0:N}$, the unknown model parameters $\bftheta$ are estimated as
\begin{equation} \label{eq:optimProblem}
    \hat{\bftheta} = \arg \min_{\bftheta} J(\bftheta, \bfz_{0:N}), 
\end{equation}
where $J:\bbR^{n_{\theta}\times (N+1)n_{z}} \mapsto \bbR^{++}$ is a criterion selected by the designer.

\section{Identification Methods} \label{sec:IdentificationMethods}
In this section two methods for estimating the model parameters are presented. For the parameter estimation methods that use the maximum likelihood approach~\cite{Tryon1983}, the criterion $J$ is a non-linear function of the model parameters $\bftheta$. Although a number of iterative optimization algorithms exist, their use can be challenging as far as the convergence and selection of starting estimate are concerned. Therefore, both proposed methods aim to avoid solving large non-linear optimization problems by using a suitable transformation of available measurements

\subsection{Method based on Allan covariance (ACOV)} \label{subsec:AVAR_Methods}
The ACOV can be defined for measurement sequences $\{z_{i,k}\}_{k=0}^{\infty}$ and $\{z_{j,k}\}_{k=0}^{\infty}$ as
\begin{multline} \label{eq:acov_meas}
   \sigma^2_{i,j} (\tau_m)  = \frac{1}{2\tau_m^2} \mean \left\{ (z_{i,k+2m} - 2z_{i,k+m} + z_{i,k}) \right.\\ \times \left. (z_{j,k+2m} - 2z_{j,k+m} + z_{j,k}) \right\},
\end{multline} 
where $\tau_m=m \Ts \in \bbR^{++}$ is an averaging time with $m \in \bbN$. The ensemble model~\eqref{eq:ensemble_state}--\eqref{eq:ensemble_meas} and the theory of stochastic processes can be employed to get an analytical expression that relates the model parameters $\bftheta$ to the ACOV $\sigma^2_{i,j} (\tau_m)$. If $i \neq j$, the ACOV is given as
\begin{multline} \label{eq:acovAnalytic1}
    \sigma^2_{i,j} (\tau_m) =  q_1^{(1)} \frac{1}{\tau_m} + q_2^{(1)} \frac{\tau_m}{3} +  r_{ij} \frac{3}{\tau_m^2} \\
    + \left( d^{(i+1)} - d^{(1)} \right) \left( d^{(j+1)} - d^{(1)} \right) \frac{\tau_m^2}{2}.
\end{multline}
If $j = i$ the ACOV reduces to AVAR and it is given as
\begin{multline} \label{eq:acovAnalytic2}
    \sigma^2_{i,i} (\tau_m) = \left( q_1^{(1)} + q_1^{(i+1)} \right) \frac{1}{\tau_m} + \left( q_2^{(1)} + q_2^{(i+1)} \right) \frac{\tau_m}{3}\\
    +  r_{ii} \frac{3}{\tau_m^2} + \left( d^{(i+1)} - d^{(1)} \right)^2 \frac{\tau_m^2}{2}.
\end{multline}

The expressions \eqref{eq:acovAnalytic1}--\eqref{eq:acovAnalytic2} show that for a fixed $\tau_m$, the ACOV is a linear function of all model parameters except for the drifts. To avoid solving a large non-linear least squares problem, the parameter estimation is split into a linear least squares problem and a smaller non-linear least squares problem by introducing auxiliary parameters $f_{ij}$ as
\begin{multline}
    f_{ij} = g \left( d^{(i+1)},d^{(j+1)},d^{(1)} \right) =\\
    \left( d^{(i+1)} - d^{(1)} \right) \left( d^{(j+1)} - d^{(1)} \right),\label{eq:f}
\end{multline}
where $g:\bbR^{3}\mapsto\bbR$ is a known function and $i=1,\ldots,n_z$, $j=i,\ldots,n_z$.

Assuming that ACOV estimates $\hat{\sigma}^2_{i,j}\in\bbR^{\ell}$ are computed using measurements $\bfz_{0:N}$ for $\ell$ different averaging times $\bar{\tau}_{p} = m_{p}\Ts$, where $m_{p}\in\{1,2,\ldots, \lfloor N/2 \rfloor \}$ for $p=1,2,\ldots,\ell$, it is possible to use \eqref{eq:acovAnalytic1}--\eqref{eq:acovAnalytic2} to write the following linear regression model
\begin{equation} \label{eq:compact}
    \bfza = \bfPhi \bfthetaa + \bfepsilon,
\end{equation}
% \todo[inline]{Discuss the linear regression model with LK.  For example $\bar{\bfvarphi}$ does not seem to fit in dimension with $\bfI_{n_z}\otimes \bfvarphi$ }
where $\bfza\in \bbR^{n_{z_a}}$  is a vector given as
\begin{multline} \label{eq:zaDef}
    \bfza = 
   \vertcat \left( \hat{\sigma}^2_{1,1}, \hat{\sigma}^2_{2,2}, \ldots, \hat{\sigma}^2_{n_z,n_z}, \right.\\
    \left.\hat{\sigma}^2_{1,2}, \hat{\sigma}^2_{1,3},\ldots, \hat{\sigma}^2_{n_z-1,n_z} \right),
\end{multline}
and its dimension is given as $n_{z_a} = n (n-1)\ell/2$.  The regression matrix $\bfPhi \in \bbR^{n_{z_a} \times n_{\theta_a}}$ is defined as
\begin{align}
\bfPhi  &=
\begin{bmatrix}
    \indfun_{n_z} \otimes \bar{\bfvarphi} & \bfI_{n_z}\otimes \bfvarphi  & \mathbf{0}_{\ell n_z \times 2n_f}  \\
    \indfun_{n_f} \otimes \bar{\bfvarphi} & \mathbf{0}_{\ell n_f \times 4n_z} & \bfI_{n_f}\otimes \bar{\bar{\bfvarphi}}   \\
\end{bmatrix}, 
\end{align}
where $n_f = n_z(n_z-1)/2$ and
\begin{align}
    \bfvarphi &= \vertcat ( \varphi_1, \varphi_2, \ldots, \varphi_\ell),& \varphi_i &= \left[ \frac{3}{\bar{\tau}_i^2}, \frac{\bar{\tau}_i^2}{2}, \frac{1}{\bar{\tau}_i}, \frac{\bar{\tau}_i}{3} \right],\\
    \bar{\bfvarphi} &=  \bfvarphi \left(
    \left[
    \begin{smallmatrix}
         0 \\  1 
    \end{smallmatrix} \right] \otimes \bfI_2 \right), & 
    \bar{\bar{\bfvarphi}} & = \bfvarphi \left(
    \left[ 
    \begin{smallmatrix}
        1 \\ 0 
    \end{smallmatrix} 
    \right] \otimes \bfI_2 \right).
\end{align} 
The vector of the unknown parameters $\bfthetaa \in \bbR^{n_{\theta_a}}$ is defined as
\begin{equation}
\begin{aligned}
\bfthetaa = [&q_1^{(1)}, q_2^{(1)}, r_{11}, f_{11}, \ldots, q_1^{(n)}, q_2^{(n)}, r_{n_z n_z}, f_{n_z n_z}, \\& r_{12}, f_{12}, \ldots, r_{n_zn_z-1} f_{n_zn_z-1} ]^{\T},
\end{aligned}
\end{equation}
and its dimension is $n_{\theta_a} = n(n+1)$.
The ACOV estimation error $\bfepsilon \in \bbR^{n_{z_a}}$ has zero mean. Since the true covariance matrix of $\bfepsilon$ is difficult to obtain, it is approximated by a diagonal covariance matrix $\bar{\bfR} \in \bbR^{n_{z_a} \times n_{z_a}}$. Each diagonal element represents the corresponding variance of ACOV estimate $\hat{\sigma}^2_{i,j} (\bar{\tau}_{p})$ that is approximately given as
\begin{equation} \label{eq:varOfAvar}
    \var \left\{\hat{\sigma}^2_{i,j} (\bar{\tau}_p) \right\}\approx 2\hat{\sigma}^2_{i,j} (\bar{\tau}_{p}) / \nu,
\end{equation} 
where $\nu\in\bbR^{++}$ is an effective degree of freedom. The value of this parameter generally varies depending on the type of noise. However, a random-walk frequency modulation noise can be chosen as a conservative option with $\nu \approx N/m_{p}$.

The estimate of the unknown parameters $\bfthetaa$ is obtained by solving the minimization problem \eqref{eq:optimProblem} with the criterion
\begin{equation} \label{eq:J2_def}
  J(\bftheta, \bfz_{0:N}) = ||\bfW^{1/2} (\bfza - \bfPhi \bftheta) ||^2, 
\end{equation}
where $\bfW \in \bbR^{n_{z_a} \times n_{z_a}}$ is a weighting matrix selected as $\bfW = \bar{\bfR}^{-1}$. To obtain unique estimates of the frequency drifts, the drift $d^{(1)}$ is assumed to be known and other drifts are estimated by solving the non-linear least squares problem
\begin{align}
    \min_{d^{(2)},\ldots,d^{(n)}} \sum_{i=1}^{n_{z}}\sum_{j=i}^{n_z} \left(\hat{f}_{ij} - g\left( d^{(i+1)},d^{(j+1)},d^{(1)} \right)\right)^{2}.
\end{align}
Finally, note that the number of averaging times $\ell$ and their particular values represent design parameters of the ACOV method.

\subsection{MDM estimation method}\label{subsec:MDM_Methods}
The unique properties of the MDM stem from the definition of the residue, which can be calculated from available
measurements and shown to be a linear function of the immeasurable state and measurement noises.

\subsubsection{Augmented Measurement Vector}
Derivation of the MDM starts with the augmented measurement vector as \cite{KoDuPuSt:26}
\begin{align}
	\bfZ_k = \calO\bfx_k + \bfGamma\bfW_k + \bfV_k \label{Z}
\end{align}
for $k=0,\ldots,N-(L\!-\!1)$, where $L\!\geq\!1$ is a user-defined parameter, $\bfZ_k\in\real^{Ln_z}$, $\bfW_k\in\real^{(L-1)n_x}$, $\bfV_k\in\real^{Ln_z}$ $\calO\in\real^{Ln_z\times n_x}$, and $\bfGamma\in\real^{Ln_z\times (L-1)n_x}$ are defined 
as 
\begin{subequations}\label{eq:matrices}
\begin{align}
    \!\!\!\bfZ_k\!&\triangleq\!\!\left[\begin{smallmatrix} \bfz_k \\ \bfz_{k+1} \\ \bfz_{k+2} \\ \vdots \\ \bfz_{k+L-1} \end{smallmatrix}\right]\!\!,\ 
    \bfW_{k}\!\triangleq\!\!\left[\begin{smallmatrix} \bfw_k \\ \bfw_{k+1} \\ \bfw_{k+2} \\ \vdots \\ \bfw_{k+L-2} \end{smallmatrix}\right]\!\!,\ 
    \bfV_k\!\triangleq\!\!\left[\begin{smallmatrix} \bfv_k \\ \bfv_{k+1} \\ \bfv_{k+2} \\ \vdots \\ \bfv_{k+L-1} \end{smallmatrix}\right]\!\!,\!\!
    \\
    \!\!\calO\!&\triangleq\!\!\left[\begin{smallmatrix} \bfH \\ \bfH\bfF \\ \bfH\bfF^2 \\ \vdots \\ \bfH\bfF^{L-1} \end{smallmatrix}\right]\!\!,\ 
    \bfGamma\!\triangleq\!\!\left[\begin{smallmatrix}
        \bfnul_{n_z\times n_x} & \bfnul_{n_z\times n_x}& \cdots & \bfnul_{n_z\times n_x} 
		\\ \bfH & \bfnul_{n_z\times n_x} & \cdots & \bfnul_{n_z\times n_x} 
		\\ \bfH\bfF & \bfH & \cdots & \bfnul_{n_z\times n_x}
		\\ \vdots & \vdots & \ddots & \vdots
		\\ \bfH\bfF^{L-2} & \bfH\bfF^{L-3} & \cdots& \bfH
    \end{smallmatrix}\right]\!\!.
\end{align}
\end{subequations}

\subsubsection{Residue Definition}
The known augmented measurement vector $\bfZ_k$ \eqref{Z} depends on the unknown state and measurement noises whose mean values and covariance matrices (CM) are sought.
To eliminate the state, define the non-zero annihilation matrix $\anni(\calO)\in\real^{\aO\times Ln_z}$ of the matrix $\calO$ such that 
\begin{align}
    \anni(\calO)\calO=\bfnul_{\aO\times n_x}.\label{eq:am}
\end{align}
where $\aO\!=\!Ln_z\!-\!\rank(\calO)$ and $\rank(\calO)$ denotes a rank of the matrix $\calO$.
Then, the residue $\bbfZ_k\!\in\!\real^{\aO}$ is defined as\linebreak\\[-7mm]
\begin{align}
	\bbfZ_k=\anni(\calO) \bfZ_k=\calA\calE_{k}, \label{mpeMDM}
\end{align}
where $\calA\triangleq \anni(\calO)
\begin{bmatrix}\bfGamma,&\bfI_{Ln_z}\end{bmatrix}\in\real^{\aO\times \re}$ vector
$\calE_{k}\triangleq\begin{bsmallmatrix}\bfW_k\\\bfV_k \end{bsmallmatrix}\in\!\real^{\re}$ 
with $\re\!=\!(L-1) n_w\!+\!Ln_v$.

\subsubsection{State Noises Mean Identification}

Consequently, the mean of the residue is a linear function of the mean vectors of the state noise (drift), i.e.,
\begin{align}
	\!\!\!\!\calR_{\bbfZ}&\triangleq\mean\!\left[\bbfZ_k\right]
    =\calA\calR_{\calE}
    =\anni(\calO)\bfGamma\calR_{\bfW}\\
   % &=\anni(\calO_{k})\bfGamma_k(\bf1_{(L-1)\times1}\otimes\begin{bmatrix}\Ts^2/2 \\ \Ts \end{bmatrix}\calR_{\bfw}\\
    &=
    \anni(\calO)\bfGamma
    \underbrace{\!
    \left(\indfun_{L-1}
    \otimes
    \!\bfI_{n}\!
    \otimes
    \!
    \begin{bmatrix} 
    \Ts^2/2 \\ \Ts 
    \end{bmatrix}
    \right)\!
    }_{\Upsilon_{\bfd^{(1,2,\cdots\!,n)}}}
    \underbrace{
    \begin{bmatrix} 
    d^{(1)}\\ d^{(2)}\\  \vdots\\ d^{(n)}
    \end{bmatrix}
    }_{\bfd^{(1,2,\cdots\!,n)}}
    \label{meanTrue}\!\!\!\!
\end{align}

The residue mean $ \calR_{\bbfZ} $ is unknown.
However, its sample-based estimate $\widehat{ \calR_{\bbfZ}}= \frac{1}{N\!-\!L\!+\!2}\!\sum_{k=0}^{N-L+1}\bbfZ_k $ is computed from measurements and \eqref{meanTrue} can be modified as
\begin{align}
	\underbrace{\widehat{ \calR_{\bbfZ}}}_{\hspace{-43px}\text{Known vector}\hspace{-20px}}
	=
\underbrace{\anni(\calO)\bfGamma\Upsilon_{\bfd^{(1,2,\cdots\!,n)}}\!}_{\hspace{-45px}\text{Known matrix}\hspace{-45px}}
	\overbrace{\bfd^{(1,2,\cdots\!,n)}\!\!}^{\hspace{-30px}\text{Sought drift vector}\hspace{-30px}}
	+
	\underbrace{\!\!\!\bfzeta\!\!\!}_{\hspace{-40px}\text{Unknown \textit{zero-mean} vector}\hspace{-40px}}.\hspace{-10px}\label{meanMeasBeta}
\end{align}
 Unfortunately, matrix $\anni(\calO)\bfGamma_k\Upsilon_{\bfd^{(1,2,\cdots\!,n)}}$ does not have full column rank. To obtain a unique estimate, the drift value of the pivot clock $d^{(1)}$ is typically assumed to be known, and the remaining drifts are estimated. The drift estimates can be obtained by modifying equation \eqref{meanMeasBeta} as follows
\begin{align}
	\widehat{\bfd^{(2,\cdots\!,n)}} \!&=\!\left(\anni(\calO)\bfGamma\Upsilon_{\bfd^{(2,\cdots\!,n)}}\right)^{\dagger}\!\left(\widehat{\calR_{\bbfZ}}-\anni(\calO)\bfGamma\Upsilon_{d^{(1)}}d^{(1)}\right)
\end{align}
where $\Upsilon_{\bfd^{(2,\cdots\!,n)}} = \indfun_{L-1} \otimes 
\!
\begin{bsmallmatrix}
    \bfnul_{1 \times (n-1)}  \\ 
    \bfI_{n-1}
\end{bsmallmatrix}
\! \otimes \!
\begin{bsmallmatrix}  
    \Ts^2/2 \\ 
    \Ts 
\end{bsmallmatrix}$, 
$\Upsilon_{d^{(1)}}= \indfun_{L-1} \otimes \!
\begin{bsmallmatrix}
    1 \\ \bfnul_{(n-1) \times 1}\\
\end{bsmallmatrix}\! \otimes \! \begin{bsmallmatrix}  \Ts^2/2 \\ \Ts \end{bsmallmatrix}$, and
$ \Upsilon_{\bfd^{(1,2,\cdots\!,n)}}=\begin{bmatrix}
    \Upsilon_{d^{(1)}} & \Upsilon_{\bfd^{(2,\cdots\!,n)}}
\end{bmatrix}$. The \textit{zero-mean} residue $ \tbfZ_k$ is defined as 
\begin{align}
	\tbfZ_k &= \bbfZ_k - \anni(\calO)\bfGamma\Upsilon_{\bfd^{(1,2,\cdots\!,n)}}
    \begin{bmatrix}
	    d^{(1)}\\ \widehat{d^{(2,\cdots\!,n)}} 
	\end{bmatrix}.
\end{align}

\subsubsection{Relation of Residue and Noises Covariances}
Such definition of the residue $\tbfZ_k$ \eqref{mpeMDM} leads to the following key properties. The residue \eqref{mpeMDM} 
i) is a \textit{linear} function of the unavailable state and measurement noises and 
ii) can be computed from the measurements.

Consequently, the residue covariance is a linear function of the CMs of the state and measurement noises, i.e.,
\begin{align}
	\calR_{\tbfZ^2}&\triangleq\mean\!\left[\tbfZ_k^{\otimes^2}\right]
    =\calA^{\otimes^2}
	\calR_{\calE^2},\label{noncentMom}
\end{align}
where $\calA^{\otimes^2}\!\triangleq\!\calA\otimes\calA$ denotes the Kronecker power, $\!\calR_{\tbfZ^2}\in\real^{\aO^2}$ is the vectorised residual \textit{covariance} containing the CM elements of the residue $\tbfZ_k$ and $\calR_{\calE^2}\in\real^{\re^2}$ contains elements of the CM of the augmented state and measurement noise vector $\calE_{k}$.
\subsubsection{Noise CM Parametrisation via Structure-Defining Matrices}
The noise covariance $\calR_{\calE^2}$ contains multiple copies of \textit{unique} elements of the noise CMs $\bfQ$ and $\bfR$, due to either the construction of the residue, application of the Kronecker algebra, or the CM symmetry. For efficient estimation of the noise CM, it is necessary to express the \textit{unique} elements of the noise CMs $\bfQ, \bfR$ from the vector $\calR_{\calE^2}$ in \eqref{noncentMom}. An appealing parametrisation, introduced in \cite{Be:74}, defines $\bfQ, \bfR$ using a set of known (or user-defined) structure-defining matrices $ \bfB_Q^{(i)}$ and $ \bfB_R^{(i)} $ and unknown (and sought) parameters $\theta_\alpha^{(i)}$ as
\begin{align}
	\bfQ = \sum_{i=1}^{n_{\theta_\alpha}}\theta_\alpha^{(i)}\bfB_Q^{(i)},\ \ \bfR = \sum_{i=1}^{n_{\theta_\alpha}}\theta_\alpha^{(i)}\bfB_R^{(i)}.
\end{align}
where $n_{\theta_\alpha}=n(n+3)/2$ and $\bftheta_\alpha$, $\bfB_Q^{(i)}$ a $\bfB_R^{(i)}$ are defined as\!\!\!\!
\begin{subequations}\label{basMat}
\begin{align}
\!\!\!\!\bftheta_\alpha \!&=\! \begin{bmatrix} q_1^{(1)}, \!\cdots\!, q_1^{(n)}, q_2^{(1)}, \!\cdots\!, q_2^{(n)}, r_{11}, r_{12}, \!\cdots\!, r_{n_zn_z}\!\end{bmatrix}\T\!\!,\hspace{-2mm}
\\
\!\!\!\!\bfB_Q^{(i)}&=
\begin{cases}
\diag\left(
\begin{bsmallmatrix}\!
    \bfnul_{(i-1)\times 1} \\ 1 \\  \bfnul_{(n-i)\times 1}
\end{bsmallmatrix}\!
\right)\!\!
\begin{bsmallmatrix} 
    \Ts & 0 \\ 
    0 & 0
\end{bsmallmatrix}, \ 1\leq i \leq n
\\[3mm]
\diag\left(
\begin{bsmallmatrix}\!
    \bfnul_{(i-1)\times 1} \\ 1 \\ \bfnul_{(n-i)\times 1}
\end{bsmallmatrix}\!
\right)\!\!
\begin{bsmallmatrix} 
    \Ts^3/3 & \Ts^2/2 \\ 
    \Ts^2/2 & \Ts 
\end{bsmallmatrix}, \ n+1\leq i \leq 2n
\\
\bfnul_{n\times n}, \ 2n+1\leq i \leq 2n+n_z(n_z+1)/2
\end{cases}
\\
\!\!\!\!\bfB_R^{(i)}&= 
\begin{cases}
\bfnul_{n\times n}, \ 1\leq i \leq 2n
\\
\calI(i-2n),  \ 2n+1\leq i \leq 2n+n_z(n_z+1)/2
\end{cases}\hspace{-5mm}
\end{align}
\end{subequations}
where $\calI(j)$ is a function that returns a zero matrix of dimension $n_z \times n_z$ containing ones only at the positions where the $i$-th unique element of the matrix $\bfR$ is located.
For example, for the first unique element $r_{11}$, the matrix is $\calI(1)=\diag(\begin{bsmallmatrix}
    1 & \bfnul_{1\times n_z-1}
\end{bsmallmatrix})$, and for the third element $r_{13}$, the matrix is $\calI(3)=\diag\!\begin{bsmallmatrix}\!\begin{bsmallmatrix}0 & 0 & 1\\0 & 0 & 0\\1 & 0 & 0 \end{bsmallmatrix} ,& \bfnul_{n_z-3\times n_z-3} \end{bsmallmatrix}\!$.
%and $\blkdiag[\cdot]$ denotes a block-diagonal matrix.

\subsection{Residue Covariance}

Then, the noise CMs parametrisation \eqref{basMat} allows to incorporate a priori knowledge of the noise structure using the \textit{minimum} number of noise CM parameters gathered in the vector $\bftheta_\alpha$. The vectors $\calR_{\calE^2}$ and $\bftheta_\alpha$ are related as follows
\begin{subequations}\label{NpsiM}
\begin{align} 
    \hspace{-8mm}&\calR_{\calE^2}\triangleq 
    \mean\!\left[\calE_k^{\otimes^2}\right]
	=\!
    \left(\diag\!\left[\bfI_{(L-1)}\!\otimes\!\bfQ,\ \bfI_{L}\!\otimes\!\bfR\right]\right)_\mathsf{V}\hspace{-2mm} 
	\\
	&=\!
	\underbrace{\!
        \Bigg[\!
		\begin{smallmatrix}\!      
            \begin{bsmallmatrix}
				\!\bfI_{(L-1)}\otimes\bfB_Q^{(1)} & \bfnul\\
				\bfnul & \bfI_{L}\otimes\bfB_R^{(1)}\!\!
			\end{bsmallmatrix}_\mathsf{\!V},
			&\!\!
			\begin{bsmallmatrix}
				\!\bfI_{(L-1)}\otimes\bfB_Q^{(2)} & \bfnul\\
				\bfnul & \bfI_{L}\otimes\bfB_R^{(2)}\!\!
			\end{bsmallmatrix}_\mathsf{\!V},\!
            & \! \ldots
		\end{smallmatrix}\!\Bigg]
	\!}_{\Upsilon_{\!\bftheta_\alpha}}
	\!\bftheta_\alpha,\hspace{-1.5mm} 
\end{align}
\end{subequations}
where the notation $(\cdot)_\mathsf{V}$ stands for vectorisation of the matrix.

\subsection{Noise Parameter Estimation}
The residue covariance $ \calR_{\tbfZ^2} $ is unknown. However, its sample-based estimate $\widehat{ \calR_{\tbfZ^2}}= \frac{1}{N\!-\!L\!+\!2}\!\sum_{k=0}^{N-L+1}\tbfZ_k^{\otimes^2} $ can be computed and equations \eqref{noncentMom} and \eqref{NpsiM} can be modified as
\begin{align}
	\underbrace{\widehat{ \calR_{\tbfZ^2}}}_{\hspace{-43px}\text{Known vector}\hspace{-20px}}
	=
   \underbrace{\calA^{\otimes^2}\Upsilon_{\bfalpha}}_{\hspace{-20px}\text{Known matrix}\hspace{-20px}}
	\overbrace{\bftheta_\alpha}^{\hspace{-25px}\text{Sought noise parameters}\hspace{-25px}}
	+
	\underbrace{\!\!\!\bfnu\!\!\!}_{\hspace{-20px}\text{Unknown \textit{zero-mean} vector}\hspace{-60px}}.\hspace{-10px}\label{CovParEst}
\end{align}
The noise parameters $\bftheta_\alpha$ can be estimated by \eqref{CovParEst} as follows
 \begin{align}
	\widehat{\bftheta_\alpha} &= \left(\calA^{\otimes^2}\Upsilon_{\bfalpha}\right)^{\!\dagger} \widehat{ \calR_{\tbfZ^2}}
\end{align}
Note that the resulting quality of the estimates i.e., the drift $\bfd^{(i)}$ and the noise parameter $\bftheta_\alpha$, is influenced by the choice of the parameter $L$ as well as the re-sampling period $\Ts$. Based on observations for this type of model, the following choice of $L=5$ and $\Ts=5000s$ appears to be reasonable.

\section{Results} \label{sec:Results}
\subsection{Simulation Scenario}
The ensemble consists of four active hydrogen maser (AHM) clocks with true parameters summarized in Table~\ref{Tab:Synth1} and the measurement noise covariance matrix is given as $\bfR = \num{1e-35}\times 
\left[\begin{smallmatrix}
 9& 6& 5 \\
 6& 8.7& 4 \\
 5 & 5& 9.5 \\
\end{smallmatrix}\right]$. Other parameters considered are: 
$\Ts=\qty{5}{\second}$, $N=\num{6.312e6}$ (i.e., this corresponds to a one-year data period)  and $\ell = 20$, where averaging times $m$ are evenly spaced in a log-space ranging from 1 to $\num{3.15e6}$. 

\begin{table}[!htbp]
    \caption{Specification of ensemble clocks parameters}
    \centering
    \begin{tblr}{
        colspec={c|cccc}         
        }
        & clk1 & clk2 & clk3 & clk4\\
        \hline          
         $q_1$ ($\times\num{1e-27}$) & 1  & 1.5 & 5 & 7 \\
         $q_2$ ($\times\num{1e-35}$) & 0.1 & 2 & 1.5 & 2.5 \\
         $d$ ($\times\num{1e-21}$)& 0 & 8 & 7.5 & 3 \\
    \end{tblr}
    \label{Tab:Synth1}
\end{table} 

\begin{figure*}[!ht]
    \centering
    \includegraphics[width=0.9\textwidth]{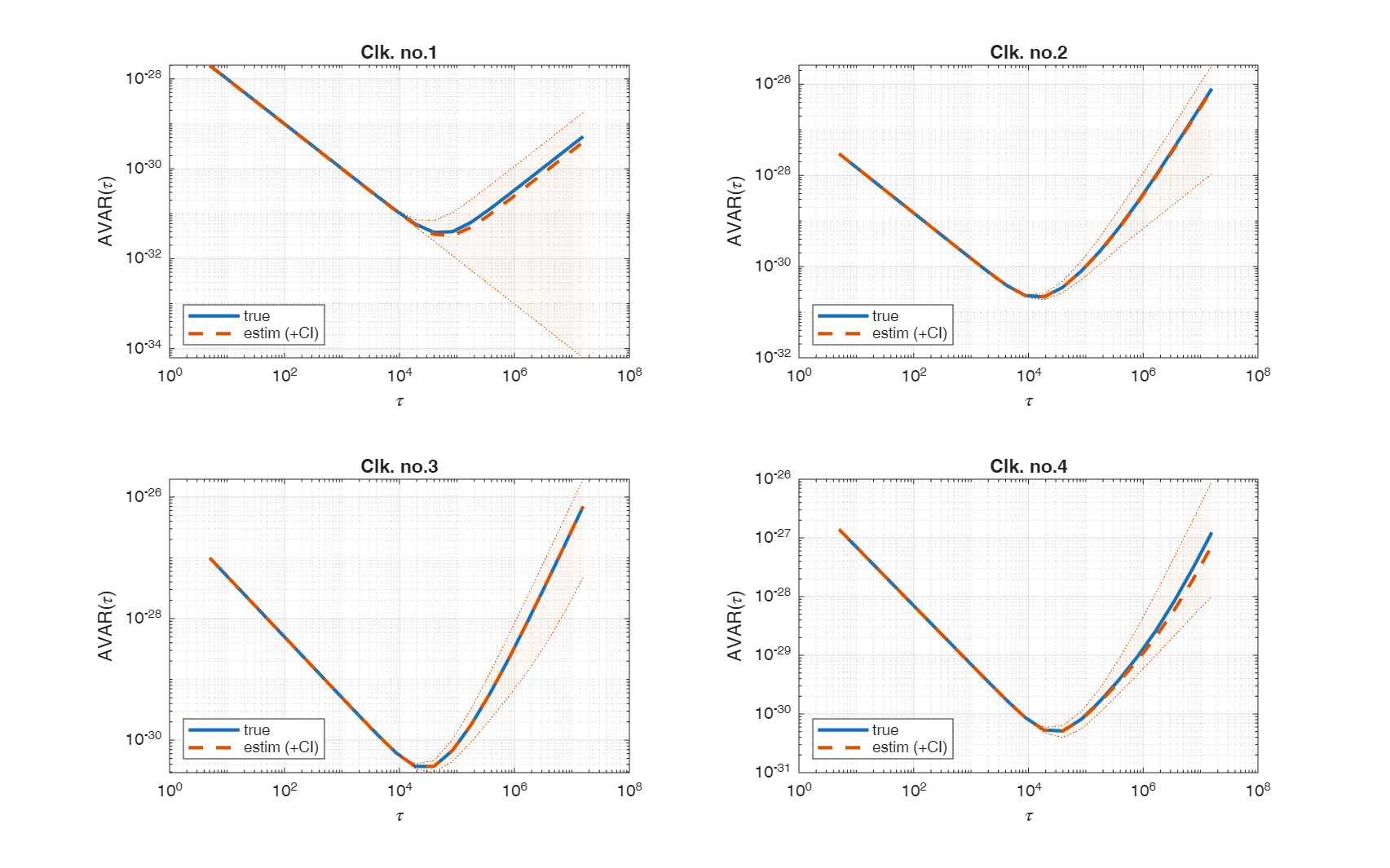}
    \caption{Clocks AVAR for true and estimated parameters (ACOV method - simulation).}
    \label{fig:clkACOV}
\end{figure*}

\begin{figure*}[!ht]
    \centering
    \includegraphics[width=0.9\textwidth]{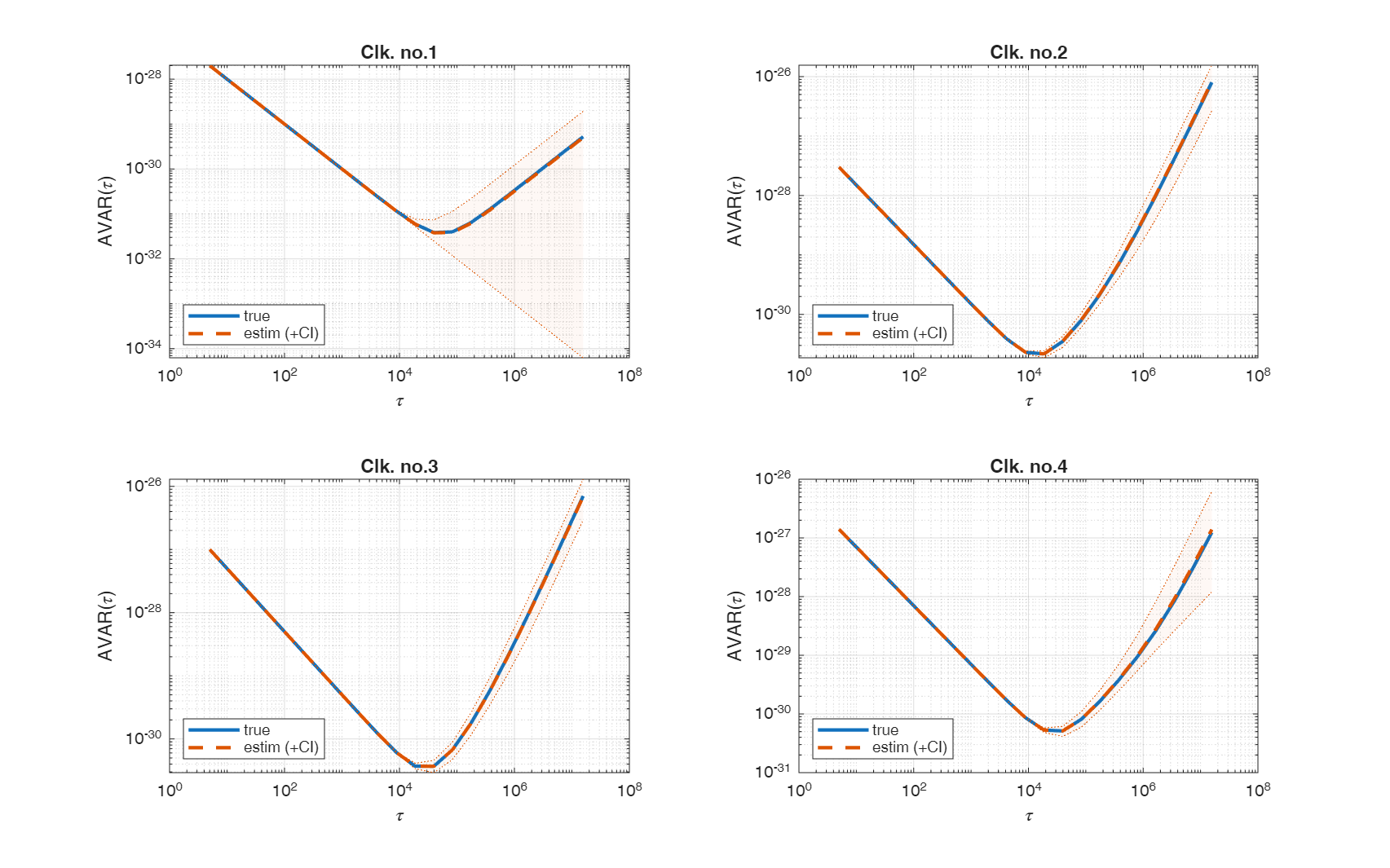}
    \caption{Clocks AVAR for true and estimated parameters (MDM method - simulation).}
    \label{fig:clkMDM}
\end{figure*}

The resulting average and standard deviations of the estimates for both methods (ACOV and MDM) based on 100 Monte Carlo (MC) simulations are summarized in Figures~\ref{fig:clkACOV} and \ref{fig:clkMDM}. The figures present the Allan variance (AVAR) of the individual clocks computed analytically using the true parameters (blue line) and using the average values of the estimated parameters from MC simulations (dashed red line). The confidence intervals of the AVAR are shown as shaded areas. The AVARs computed using MC average of parameter estimates are close to the AVARs computed using true parameters. The MDM method seems to provide parameter estimates of slightly better quality. In both cases, a wide confidence interval is obtained for the pivot clock for averaging times greater then \num{1e4}. This behavior can be attributed to (i) the assumption of zero drift for the pivot clock, (ii) the low value of the parameter $q_2$ for the pivot clock, and (iii) the limited length of the measurement record. Although the width of a confidence intervals could be further decreased by increasing the duration of measurement, a substantial reduction would require measurement duration on the order of decades.

\subsection{Real-data}
Tthe historical records of measurements were provided by the ESTEC timing laboratory during year 2022 as part of the ESA project NAVISP-EL1-056. The ensemble consists of three AHM atomic clocks iMaser3000 \cite{SAFRAN2022:iMaserDatasheet}. The nominal sampling period of the phase difference measurements is \qty{5}{\second} and the dataset contains \num{1.59e6} sampling instants (i.e., this corresponds to a three-months data period). The averaging times $m$ are evenly spaced in a log-space ranging from 1 to $\num{7.95E5}$, where $\ell = 20$. 

\begin{figure}[!ht]
    \centering
    \includegraphics[width=0.48\textwidth]{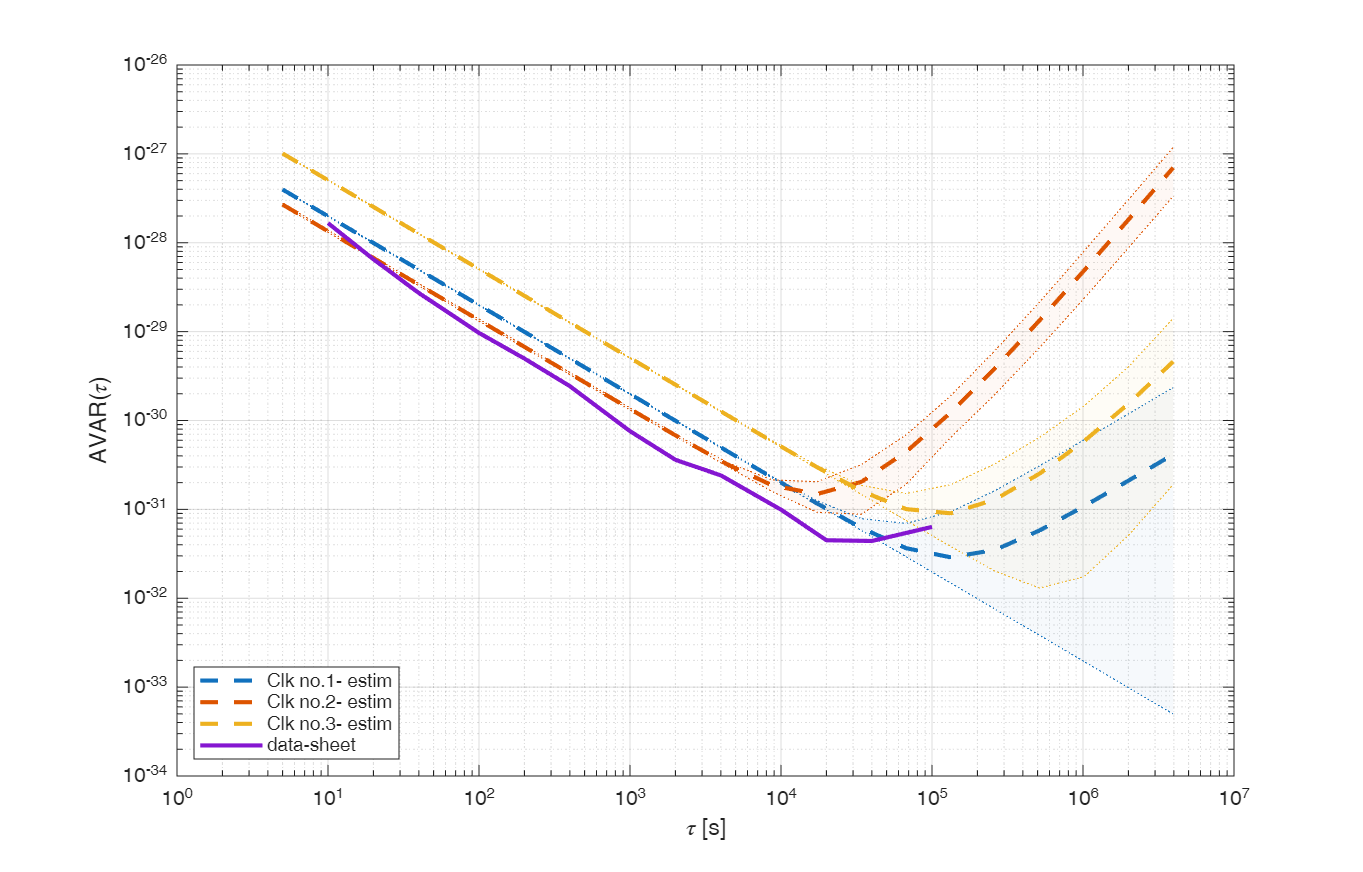}
    \vspace{-0.5cm}
    \caption{AVAR of the iMaser3000 atomic clock adopted from a datasheet and AVAR of estimated clocks (ACOV method - real data).}
    \label{fig:clkACOVreal}
\end{figure}

\begin{figure}[!ht]
    \centering
    \includegraphics[width=0.48\textwidth]{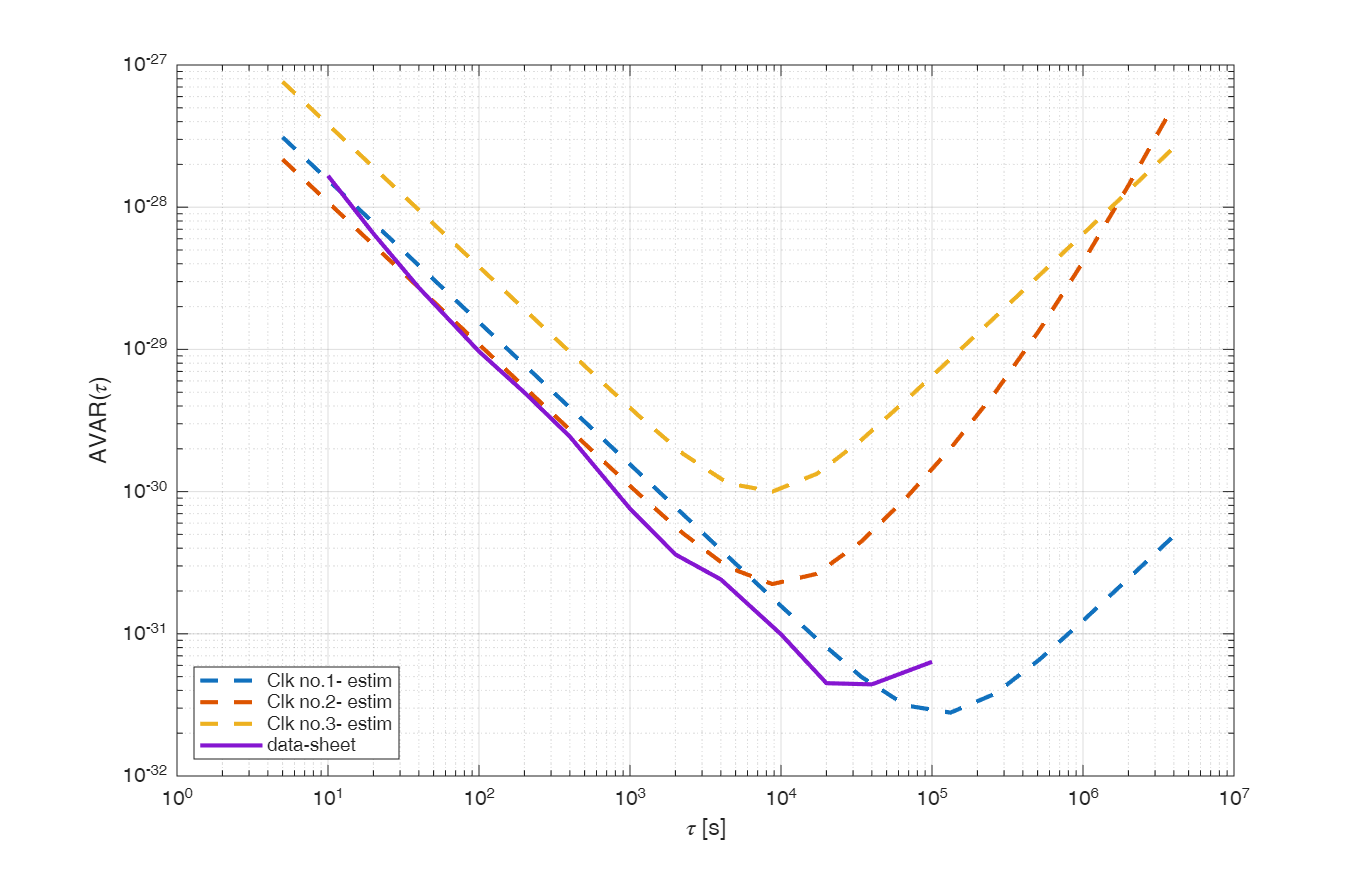}
    \vspace{-0.5cm}
    \caption{AVAR of the iMaser3000 atomic clock adopted from a datasheet and AVAR of estimated clocks (MDM method - real data).}
    \label{fig:clkMDMreal}
\end{figure}

The obvious measurement outliers were removed from the real measurements before the parameter estimation methods were applied. The AVARs of individual clocks computed using estimated parameters (dashed lines) and AVAR obtained from data-sheet (solid line) are depicted in Figures~\ref{fig:clkACOVreal} and \ref{fig:clkMDMreal}. Contrary to the simulation scenario, the parameter estimates obtained by the proposed methods show significant discrepancies. For clocks 1 and 2, the parameter estimates differ by up to \qty{20}{\percent}, whereas for clock 3 the differences span several orders of magnitude. These discrepancies may be attributed to (i) an insufficient amount of data (the available three-month record may be too short, leading to high estimation uncertainty) (ii) the presence of additional noise types in real measurements that are included in the model, and (iii) anomalies in measurements that were not fully removed during preprocessing. However, the results provide an indication of the achievable estimation accuracy under realistic conditions. This insight is particularly relevant for subsequent use of the ensemble model in combination and detection algorithms, which should be designed to remain robust in the presence of such uncertainties.

\section{Conclusions} \label{sec:Conclusions}
The article presents two identification methods for determining ensemble clocks noise parameters. The performance was evaluated on both synthetic and real data. While the accuracy of the parameter estimation is primarily constrained by sample size, both methods prove to be computationally efficient, ensuring reliable characterization of of noise characteristics under realistic conditions.

\bibliographystyle{IEEEtran}
\bibliography{literatura_ivo}

@inproceedings{Barnes1983a,
    author = {Barnes, James A. and Jones, Richard H. and Tryon, Peter V. and Allan, David W.},
    title = {{Stochastic models for atomic clocks}},
    booktitle = {Proceedings of the 14th Annual Precise Time and Time Interval (PTTI) Applications Planning Meeting},
    year = {1983},
    pages = {295--306}
}

@article{Tryon1983,
    author = {Tryon, Peter V. and Jones, Richard H.},
    title = {{Estimation of Parameters in Models for Cesium Beam Atomic Clocks}},
    journal = {JOURNAL OF RESEARCH of the National Bureau of Standards},
    year = {1983},
    vol = {88},
    issue = {1},
    pages = {3--16}
}

@inproceedings{Vernotte2016,
    author = {Vernotte, Fran\c{c}ois and Calosso, Claudio Eligio and Rubiola, Enrico},
    title = {{Three-Cornered Hat versus Allan Covariance}},
    booktitle = {2016 IEEE International Frequency Control Symposium (IFCS)},
    year = {2016},
    pages = {282--287},
    address = {New Orleans, LA, USA}
}

@article{Premoli1993:IEEETIM,
    author = {Premoli, Amedeo and Tavella, Patrizia},
    title = {{A Revisited Three-Cornered Hat Method for Estimating Frequency Standard Instability}},
    journal = {IEEE Transactions on Instrumentation and Measurement},
    year = {1993},
    vol = {42},
    issue = {1},
    pages = {7--13}
}

@article{Tavella1993:Metrologia,
    author = {Tavella, Patrizia and Premoli, Amedeo},
    title = {{Estimating the Instabilities of N Clocks by Measuring Differences of their Readings}},
    journal = {Metrologia},
    year = {1993},
    vol = {30},
    issue = {5},
    pages = {479--486}
}

@misc{SAFRAN2022:iMaserDatasheet,
  author       = {{SAFRAN}},
  title        = {{iMaser3000 - Product Information}},
  howpublished = {\url{https://safran-navigation-timing.com/product/imaser-3000/}},
  note         = {Online: (Available 01.04.2026)},
  year         = {2026}
}

@ARTICLE{Galleani2010:IEEECon,
  author={Galleani, Lorenzo and Tavella, Patrizia},
  journal={IEEE Control Systems Magazine}, 
  title={{Time and the Kalman Filter}}, 
  year={2010},
  volume={30},
  number={2},
  pages={44-65},
  doi={10.1109/MCS.2009.935568}}

@misc{Allan1997,
    author = {Allan, David W. and Ashby, Neil and Hodge, Clifford C.},
    title = {{The Science of Timekeeping}},
    publisher = {Hewlett Packard},
    year = {1997}
}

@ARTICLE{KoDuPuSt:26,
  author={Kost, Oliver and Duník, Jindřich and Punčochař, Ivo and Straka, Ondřej},
  journal={IEEE Transactions on Automatic Control}, 
  title={Unobservable Systems: No Problem for Noise Identification}, 
  year={2026},
  volume={71},
  number={2},
  pages={1223-1230},
  doi={10.1109/TAC.2025.3607601}}

@article{Be:74,
  title = {Estimation of Noise Covariance Matrices for a Linear Time-Varying Stochastic Process},
  author = {B{\'e}langer, P. R.},
  pages = {267--275},
  year = 1974,
  journal = {Automatica},
  number = 3,
  volume = 10
}

\end{document}